\documentclass[12pt]{article}
\usepackage{graphicx}
\usepackage[dvipsnames]{color}

\newcommand{\eps}{\varepsilon}

\newcommand{\rr}{\mathrm}
\def\be{\begin{equation}}
\def\ee{\end{equation}}
\def\ba{\begin{eqnarray}}
\def\ea{\end{eqnarray}}
\def\de{\partial}

\def\ltsima{$\; \buildrel < \over \sim \;$}
\def\simlt{\lower.5ex\hbox{\ltsima}}
\def\gtsima{$\; \buildrel > \over \sim \;$}
\def\simgt{\lower.5ex\hbox{\gtsima}}
\begin{document}
\title{\bf Mirror instability in a plasma with cold gyrating dust particles}
\author{V. V. Prudskikh\thanks{slavadhb@mail.ru},~
        L. V. Kostyukova,~ 
        Yu. A. Shchekinov\thanks{yus@sfedu.ru}\\
        Department of Physics, Southern Federal University,\\
        Rostov on Don, 344090 Russia}

\date{}
\maketitle

\begin{abstract}

In this work linear stability analysis of a magnetized dusty plasma 
with an anisotropic dust component having transversal motions much 
stronger than motions parallel to the external magnetic field, and isotropic light plasma components is described. Such a situation 
presumably establishes in a shock compressed space dusty plasma 
downstream the shock front. Oblique low-frequency magneto-hydrodynamic 
waves ($\omega\ll \omega_{cd}$, $\omega_{cd}$ being the dust cyclotron 
frequency) are shown to be undergone to the mirror instability. Consequences for nonthermal dust destruction behind shock fronts in 
the interstellar medium are discussed. 


%
%
%


\end{abstract}

\section*{I. INTRODUCTION}

As shown in \cite{rao,ver,mamun,tsyt,mendis}  the impurity of charged dust particles changes dynamical behavior of a plasma in the 
low-frequency limit where heavy dust particles dominate dispersion properties. Formally, the presence of an additional component increases the order of the dispersion equation and 
thus brings a new wave mode. One of interesting phenomena connected 
with the presence of dust in plasma is that due to a high inertia of dust grains such complex plasma when brought once into an anisotropic state, 
can then be kept anisotropic much longer than an ion-electron plasma, 
and therefore can develop specific anisotropic instabilities. Such conditions are easily established in a magnetized dusty plasma subjected to shock waves. Indeed, after crossing a shock front 
with parallel magnetic field a dust particle starts gyrating with the velocity $v_{d\perp}=3v_s/4$ relative to the light components (electrons and ions) (see discussion e.g. in \cite{mc,drain}), where $u_s$ is the shock velocity. The longitudinal 
velocity component of dust particles remains however equal to the velocity dispersion of dust before the shock $v_{d||}=v_{d0}\ll v_s$. As a rule, in 
these conditions the dust gyration period $2\pi/\omega_{cd}$ is much shorter than the dust drag time $\tau_d$: $\omega_{cd}\tau_d\gg 1$. For instance, in the interstellar dusty plasma behind a supernova shock 
with $v_s=100$ km s$^{-1}$ 

\be
\omega_{cd}\tau_d\sim {30|Z|\over n},
\ee
where the drag was assumed to be due to direct 
ion-dust collisions as the Coulomb cross-section at these 
conditions is much smaller than the kinetic one, $n$ is the plasma (ion) 
density, $Z$ is the grain charge: $|Z|\simeq 3akT/e^2$ -- it 
can be as high 
as $10^3$ behind a shock with $v_s=100$ km s$^{-1}$; here $\omega_{cd}=|Z|eB/m_dc$, $\tau_d^{-1}=m_pn\sigma_dv/m_d$, $m_d$, $m_p$ are the dust grain and the proton mass. Unless specified, we 
assume through the paper the characteristic size of dust grains of $a=0.1\mu$m, though our conclusions are weakly sensitive to the actual value of $a$ since we consider the frequency range lower than the dust cyclotron frequency in the whole range of grain sizes. In particular, as $\omega_{cd}\tau_d \propto a^{-1}$ the condition $\omega_{cd}\tau_d\gg 1$ remains valid for the whole dust size spectrum ranging from $a=10 \AA$ to 
$a=1\mu$m (for the interstellar dust properties see the review \cite{drain}). It is therefore clear that 
the instability connected with the anisotropy of the dust component can 
develop succesfully on times $t\simgt \omega_{cd}^{-1}$ provided the 
conditions for the instability are fulfilled. In this paper 
we shall describe an anisotropic instability in a plasma with isotropic electrons and ions 
of $T_e\sim T_i$, and anisotropic dust component with the transversal orbiting velocities much higher than the longitudinal velocities $v_{d\perp}\gg v_{d||}$. 

The paper is organized in the following fashion. In Sec.~II we derive the dispersion equation for magneto-hydrodynamic (MHD) waves in a three-fluid plasma; 
in Sec.~III we analyse solutions of the disperion relation and show that 
low-frequency waves are unstable; in Sec.~IV consequences for dust destruction behind shock waves in the interstellar medium are discussed. Finally we close our paper with a short summary in Sec.~V.

\section*{II. DISPERSION EQUATION}


Let us consider MHD waves in a plasma with cold dust particles gyrating 
with equal transversal velocities in an external magnetic field ${\bf B}_0=(0,0,B_0)$, in which case their distribution function is $f({\bf v})\propto 
\delta(v_z)\delta(v_\perp-v_0)$, where $v_0$ is the transversal velocity 
of dust particles relative to the plasma component: $v_0$ $v_0=3v_s/4$  for a plasma behind the shock wave moving with the velocity $v_s$; the wavevector is assumed to have the components $\bf{k}=(k_\perp,0,k_{||})$. Let us assume the phase velocity to fulfil the condition 

\begin{equation}\label{con1}
v_{Ti}\ll\frac{\omega}{k_{||}}\ll v_{Te},
\end{equation}
where $v_{Te}$ and $v_{Ti}$ are thermal velocities of the electrons 
and ions respectively. In the low-frequency limit $\omega\ll\omega_{ci}$, 
where $\omega_{ci}$ is the ion cyclotron frequency, the equation of MHD waves is \cite{krall} 

\begin{equation}\label{eq2}
\frac{k^2c^2}{\omega^2}=\eps_{yy}-\frac{\eps^2_{yz}}{\eps_{zz}}, 
\end{equation}
here $\eps_{yy}, \eps_{yz}, \eps_{zz}$ are the corresponding components 
of the permeability. Under the validity of Eq. (\ref{con1}) and the condition that the ion gyration radius is smaller than the wavelength 
$k_\perp v_{Ti}\ll\omega_{ci}$ the components have the form 

\begin{eqnarray}\label{eq3} 
\eps_{yy}=\eps_{yy}^i+\eps_{yy}^d,\hspace{1cm}\eps_{yy}^i=\frac{\omega_{pi}^2}{\omega_{ci}^2},\nonumber\\
\eps_{yz}=\eps_{yz}^e=-\frac{\omega_{pe}^2}{\omega\omega_{ce}}\frac{k_\perp}{k_{||}},\\
\eps_{zz}=\eps_{zz}^{e,i}=-\frac{\omega_{pi}^2}{\omega^2}+\frac{\omega^2_{pi}}{k_{||}^2c_s^2},\nonumber
\end{eqnarray}
where $\omega_{pe}$ and $\omega_{pi}$ are the plasma frequencies of the 
electrons and ions, $\omega_{ce}$ is the cyclotron frequency of the 
electrons, $c_s=\sqrt{T_e/m_i}$ is the ion sound speed; the contribution of dust into the permeability 
components  $\eps_{yz}$ and $\eps_{zz}$ can be shown to be negligible. 

Let us assume the flux of cold dust particles behind the shock 
front to be described by the distribution function 

\begin{equation}
f(\mathbf{v})=\frac{n_{d0}}{v_0}\delta(v_z)\delta(v_\perp-v_0),
\end{equation}
where $n_{d0}$ is the dust component density, $v_0=3v_s/4$ as defined above; the longitudinal motions are thus explicitly considered to be negligible as compared to the shock 
propagation. This distribution function corresponds therefore 
to the extreme anisotropy, like if the temperatures ratio was 
$T_{d\perp}/T_{d||}=\infty$ for maxwellian dust distributions, and 
from general point of view in these conditions anisotropic instabilities 
are always expected to grow. The component $\eps_{yy}^d$ has the form 

\begin{equation}\label{eq5}
\eps_{yy}^d=\frac{\omega_{pd}^2}{\omega}\sum_{n=-\infty}^{n=\infty}\int\limits_{-\infty}^{\infty}\frac{dv_z}{\omega-n\omega_{cd}-k_{||}v_z}\int dv_\perp v^3J^{'2}_n(x)F(v),
\end{equation}
where 

\begin{equation}
F(v)=\frac{1}{v_0v_\perp}\left[\left(1-\frac{k_{||}v_z}{\omega}\right)\frac{\de\delta(v_\perp-v_0)}{\de v_\perp}\delta(v_z)+\frac{k_{||}v_\perp}{\omega}\frac{\de\delta(v_z)}{\de v_z}\delta(v_\perp-v_0)\right],
\end{equation}
here $x=k_\perp v_\perp/\omega_{cd}$, $\omega_{pd}$ is the dust plasma 
frequency, $\omega_{cd}$, the dust cyclotron frequency.

After simple algebra in Eq. (\ref{eq5}) one arrives at 

\begin{equation}\label{eq7} 
\eps_{yy}^d=-\frac{\omega_{pd}^2}{\omega^2}\left[1+\sum_{n=-\infty}^{n=\infty}\left(\frac{k_{||}^2v_0^2J_n^{'2}(x)}{(\omega-n\omega_{cd})^2}+\frac{2\omega_{cd}^2}{k_\perp v_0}n(n^2-x^2)\frac{J_n(x)J_n'(x)}{\omega-n\omega_{cd}}\right)\right].
\end{equation}
Let us consider low-frequency electromagnetic waves with  $\omega\ll\omega_{cd}$, and the transversal wavelength significantly longer than the gyration radius of dust particles: $k_\perp v_\perp/\omega_{cd}\ll 1$. Then Eq. (\ref{eq7}) can be rewritten in a simple form  

\begin{equation}\label{eq8}
\eps_{yy}^d=\frac{\omega_{pd}^2}{\omega_{cd}^2}\left[1-\frac{k_\perp^2v_0^2}{\omega^2}-\frac{k_{||}^2v_0^2}{2\omega^2}\left(1+\frac{k_\perp^2v_0^2}{2\omega^2}\right)\right].
\end{equation}

The dispersion equation (\ref{eq2}), with accounting (\ref{eq3}) and (\ref{eq8}) can be written as 

\begin{equation}\label{eq9}
\frac{k^2c_A^2}{\omega^2}=\frac{\omega^2-k^2c_s^2}{\omega^2-k_{||}^2c_s^2}+\delta\left[1-\frac{k_\perp^2v_0^2}{\omega^2}-\frac{k_{||}^2v_0^2}{2\omega^2}\left(1+\frac{k_\perp^2v_0^2}{2\omega^2}\right)\right],
\end{equation}
where the notations $c_A^2=B_0^2/4\pi\rho_i$, $\delta=\rho_i/\rho_d$ are 
introduced, $\rho_i, \rho_d$ are the mass densities of the ions and dust.

Equation (\ref{eq9}) is bicubic with respect to $\omega$, and can be rewritten as 

\begin{eqnarray}\label{eq10}
2(1+\delta)\Omega^6-\biggl[2+\beta+2\delta f\sin^2\theta+\delta(\beta+f)\cos^2\theta\biggr]\Omega^4+\nonumber\\
\biggl[\beta+\frac{1}{2}\delta\beta f(1+\sin^2\theta)-\frac{1}{2}\delta f^2\sin^2\theta\biggr]\cos^2\theta\Omega^2+\frac{1}{4}\delta\beta f^2\sin^2\theta\cos^4\theta=0.
\end{eqnarray}
Here $\theta$ is the angle between the wavevector ${\bf k}$ and the 
external magnetic field ${\bf B}_0$, $\Omega=\omega/kc_A$, $\beta=2c_s^2/c_A^2$, $f=v_0^2/c_A^2$.

\section*{III. RESULTS}

In the lack of dust Eq. (\ref{eq10}) describes two low-frequency 
electromagnetic modes present in a warm magnetized plasma: 
the fast and the slow magnetosound waves. Indeed, when $\delta=0$, Eq. (\ref{eq10}) 
degenerates and converges to a biquadratic one with the two solutions corresponding to the fast and slow magnetosound waves. When  $\delta\not=0$ and $f=0$, Eq. (\ref{eq10}) accounts the presence of dust 
in the fast and slow waves in the limit $\omega\ll\omega_{cd}$: the solutions are the fast and the slow waves loaded by the dust with the 
factor $(1+\delta)$. For 
gyrating dust particles ($f\not=0$) the order of the dispersion equation 
increases, and a third low-frequency mode emerges 
caused by the interaction of 
gyrating dust particles with the magnetic field of the wave.  

It is readily seen that this low-frequency mode is always aperiodically unstable, and corresponds to the mirror instability described in 
\cite{rud,taji,hase}. 
A typical value of the dust-to-gas mass ratio $\delta$ in the interstellar medium is much lower than one (e.g. \cite{drain}), and as a rule $\beta$ behind the shock 
front varies around one. At the same time $f$ is of the order of the ratio of the shock velocity to the Alfv\'en speed and can significantly exceed one. In the limit $f\to \infty$ the growth rate of the unstable 
low-frequency mode can be readily estimated asymptotically as $\Gamma\sim 
f^{1/2}$, while the frequences of the wave modes are 

\be
\Omega_1^2={f\delta\over 4(1+\delta)}
\biggl[1+\sin^2\theta+\sqrt{(1+\sin^2\theta)^2+{\delta\over 1+\delta}
\sin^22\theta}\biggr]
\ee
\be
\Omega_2^2={\beta\over 2}\cos^2\theta. 
\ee

In the opposite limit $f\to 0$ the solutions are  

\be
\Gamma\simeq {f\sqrt{\delta}\over 4}\sin 2\theta,
\ee
for the growing mode, and 

\be
\Omega^2={1\over 4(1+\delta)}\biggl[2+\beta\pm
\sqrt{(2+\beta)^2-8\beta(1+\delta)\cos^2\theta}\biggr]+O(f),
\ee
for the fast and slow modes correspondingly. 

The effect of dust gyration on magnetosonic waves is illustrated on Figures 1: Fig. 1a and 1b show the frequency of the fast wave (a), and the growth rate of the third (low-frequency) mode (b) 
versus  $f$; the slow magnetosonic 
mode is rather unsensitive to $f$. Slowly gyrating and nongyrating dust particles ($f<1$) lead to a decrease of the phase velocity, which can be readely understood as due to a higher density of a dusty plasma than a dust-free plasma. 

As $f$ grows  the phase velocity increases, which is due to an enhancement of the transversal elasticity by the gyrating dust particles. Indeed, the transversal dust pressure can be defined as 

\begin{equation}
P_\perp=m_d\int v_\perp^2f(\rr{v})d\rr{v}=n_{d0}m_dv_0^2.
\end{equation}
For strong shock waves, which are harmful for dust particles, $v_0\gg c_A$, and the 
transversal pressure of dust component $\rho_d v_0^2$ becomes comparable 
(or exceeds) to the magnetic pressure  $\rho_i c_A^2/2$ even though  
$\delta=\rho_d/\rho_i\ll 1$.  

Fig. 2 shows the dependence of the frequency $\Omega$ and the growth rate $\Gamma$ on plasma-$\beta$. The phase velocity of the fast wave 
increases with $\beta$ due to a joint action of the magnetic and thermal pressure. At small $\beta$ the slow wave does not reveal significant deviations from the slow wave in an electron-ion plasma; deviations becomes 
considerable at sufficiently large $\beta$. The growth rate of the unstable mode is rather unsensitive to the effects of the plasma 
thermal pressure.  
 
The mechanisms of the instability is similar to the mechanism of the mirror instability of the slow MHD wave in a plasma with anisotropic 
ion temperature $T_{i\perp}>T_{i||}$. In our case, however, the 
isotropy of 
the electron and ion temperatures stabilize the slow and 
the fast waves, while on the other side the presence of 
a third cold gyrating dust component with an anisoptopic velocity distribution results in a new unstable mode. As expected the maximum of the growth rate $\Gamma$ is reached when the 
angle between the wave vector and magnetic field is $\pi/4$.

\noindent 

\section*{IV. DISCUSSION} 

We have shown in this paper that the presence of an anisotropic dust in 
a plasma results in development of the mirror instability in the low-frequency limit $\omega\ll\omega_{cd}$. The instability can be important for the flows behind magnetized shock waves in the interstellar medium: here dust particles keep significant transversal component during a long time $\omega_{cd} t\gg 1$ after 
crossing the shock front, and thus support development of the instability. The time of isotropization of dust grain velocities is of the order of 
$\tau_i\sim m_d/nm_p\sigma_dv_0$, where $m_p$ is the proton mass, $m_d\sim 10^{10}m_p$ is the grain mass, $\sigma_d\sim 10^{-10}$ cm$^2$, the grain cross-section, $n$ is plasma density; this gives $\tau_i\sim 10^{13}/n$ s  for typical value of the shock velocity $v_0\sim 100$ km s$^{-1}$,
thus resulting in $\omega_{cd}\tau_i\sim 10^7/n\gg 1$. It is worth noting that 
when direct collisions between protons and dust grains dominate, $\omega_{cd}\tau_i$ increases when the grain radius $a$ decreases, such that the dust component remains anisotropic on the growth time of the instability for the whole range of grain sizes. Grain-grain collisions remains unimportant in 
isotropization of the transversal dust velocities: indeed, the characteristic time of grain-grain collisions is of the order 
$\tau_{gg}\sim 1/n_d\sigma_dv_{0d}$, such that $\tau_{gg}\sim \rho\tau_i/\rho_d\sim 100 \tau_l$. Note that Coulomb collisions 
between charged dust particles are inefficient because the cross-section $\sigma_{\rm C}\sim Z^2e^4\ln\Lambda/(m_dv_d^2)^2
\sim 10^{-15}\ln\Lambda$ cm$^2$. In principle, fluctuations of dust charge can enhance $\sigma_{\rm C}$ (see, e.g. discussion in 
Ref. \cite{ivlev,ivl}), however, they can be important only for dust grains with the charge $Z\sim 1$, while in the conditions with the charge $Z\simgt 100$ what we are actually interested in, its fluctuations cannot significantly enhance $\sigma_{\rm C}$.  

As the normalized growth rate $\Gamma$ depends only on $\delta$, $\beta$, 
$f$ and the angle $\theta$, the instability
grows on a nearly Alfv\'en crossing time $\simeq (0.3-0.5)\times\lambda/c_A$ (see, Fig. 1b and 2c). The 
upper limit of the wavelength for which isotropization can become 
important, and therefore the driving mechanism weakens and the growth rate drops, 
is determined from the condition $\Gamma kc_A\tau_i<1$, and is 
of the order $\lambda\sim 10^{20}c_A/nv_0$ cm. For the conditions 
typical behind the supernovae shocks in the 
interstellar medium $c_A\sim 30$ km s$^{-1}$, and the shock velocity 
$v_0\sim 100$ km s$^{-1}$, this saturation wavelength is $\lambda\sim 3\times 10^{19}$ cm. This value is approximately 3 times of the thicknes of the shell of a typical supernova remnant when its 
expansion velocity is $v_s=100$ km s$^{-1}$.  

This means that the mirror instability induced by the anisotropy of 
dust velocities develops on all scales behind the shock front of a  supernova remnant. As the growth rate decreases with the wavelength, 
the most fast growing perturbations are expected to lie in the lower end 
of our approximation of the low-frequency limit, i.e. on several dust gyration radii. It is shown in Ref. \cite{kive} that on nonlinear stages the growing instability can form holes in the background magnetic field of several dust gyroradius in size, as observed in the experiment \cite{burl}. 

From this point of view it can have great importance for dust destruction behind strong shock waves in the interstellar medium. 
It is commonly known that strong 
($v_s>100$ km s$^{-1}$) radiative supernovae shocks efficiently destroy 
dust grains via sputtering (see e.g. \cite{drain77,mckee}). 
The destruction efficacy 
is mostly supported by betatron acceleration of the dust transversal 
motions $p_\perp^2/B(t)$=const through the growing magnetic field in a 
radiatively cooling and compressing plasma: $B\propto \rho$; $p_\perp$ is the transversal momentum of a grain. This means 
that the destruction mechanism works efficiently only behind the fronts with a 
planar magnetic field. In the flows where the  mirror instability 
operates and violates regular structure of magnetic field, the growing magnetic holes suppress acceleration of dust grains and therefore their destruction can be reduced. The exact value of the destruction rate in such conditions depends on many factors, such as the number of magnetic holes, their sizes and magnetic field strength, their lifetime, and can be only estimated in nonlinear numerical consideration. One should also mention that dust charge fluctuations \cite{ivlev,ivl} can result in acceleration/decceleration of dust particles, which in turn can affect the mirror instability. However, such effects apparently are important only when dust charge is small, close to $Z\sim 1$. In the conditions behind strong shock waves with highly charged dust grains $|Z|\sim 10^2-10^3$ the increase of $\eps_{yy}^d$ due to charge fluctuations, and therefore their  
influence on the mirror instability, seems to be small. 

\section*{IV. SUMMARY}

In this paper we have analyzed stability conditions of a dusty plasma with isotropic light components (ions and electrons) and anisotropic dust particles, with transversal velocities much higher than velocities along the external magnetic field. We have shown that 

\begin{itemize}

\item{} oblique low-frequency ($\omega\ll \omega_{cd}$) magneto-hydrodynamic waves are unstable against the mirror instability; 

\item{} this instability can be important in the interstellar dusty plasma behind strong shock waves from supernovae explosions, where the pressure of the gyrating dust particles can be comparable to the magnetic pressure behind the shock front. In these conditions the instability develops practically in the whole range of spatial scales, though the most fast growing wavelengths correspond to the lengths of several dust gyration radii; 

\item{} the instability operates to destroy regular structure of the magnetic field behind the shock front, and can reduce the destruction efficiency of the interstellar dust by shock waves. 

\end{itemize}

\section*{Acknowledgements}

We acknowledge critical remarks by the anonymous referees. This work was supported in part by the Federal Agency of Science and  Innovations (project code 02.740.11.0147), by RFBR (project codes  08-02-00933 and 09-02-97019) and by the German Science Foundation (DFG) 
through the Sonderforschungsbereich 591.

\newpage

\begin{figure}\center
\includegraphics[height=6.0cm]{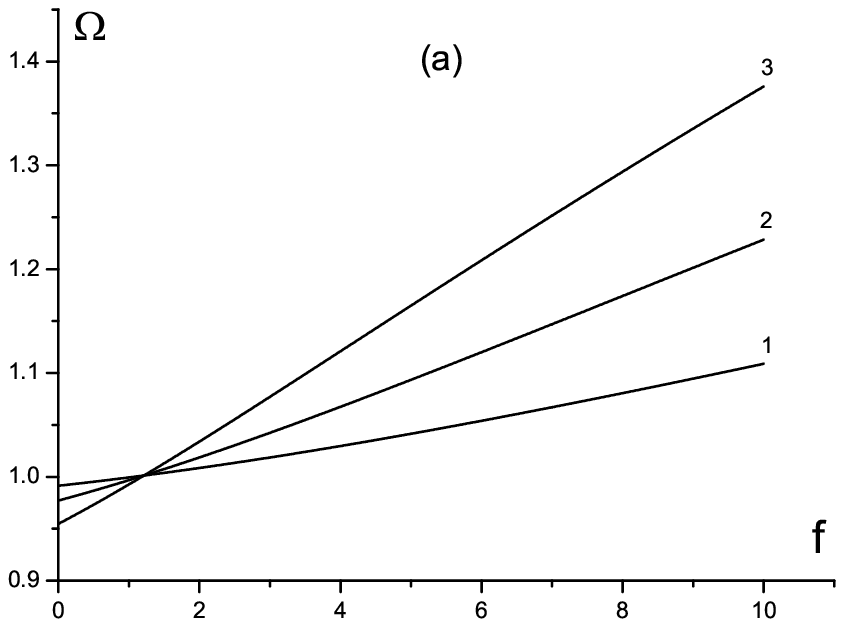}
\includegraphics[height=6.0cm]{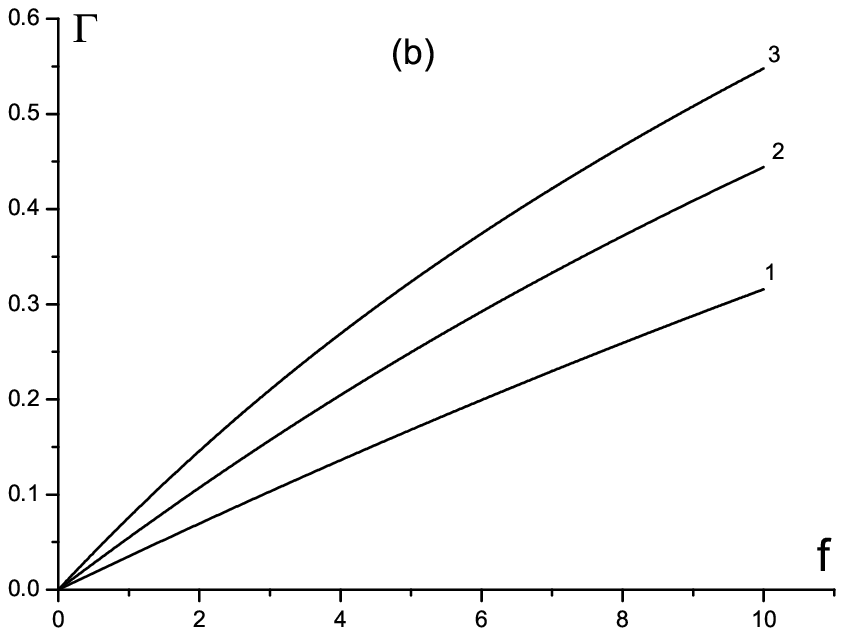}
\caption{ $\Omega$ of the fast MHD wave (a) and $\Gamma$ of the growing low-frequency mode (b) versus the gyration rate  
$f$: $\delta=0.02,~0.05,~0.1$ from bottom to top;  $\beta=0.01$, $\rr{cos}^2\theta=0.5$.}
\end{figure} 

\begin{figure}\center
\includegraphics[height=6.0cm]{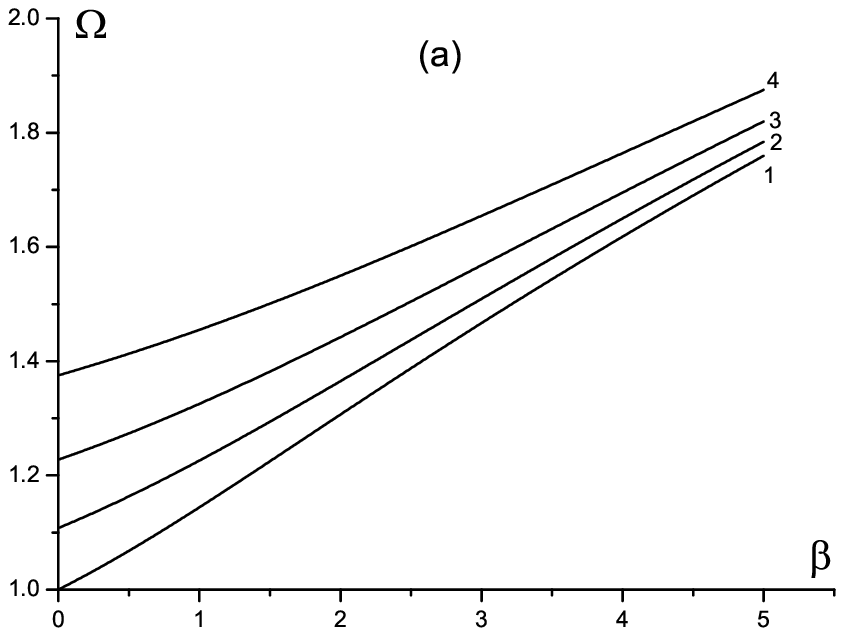}
\includegraphics[height=6.0cm]{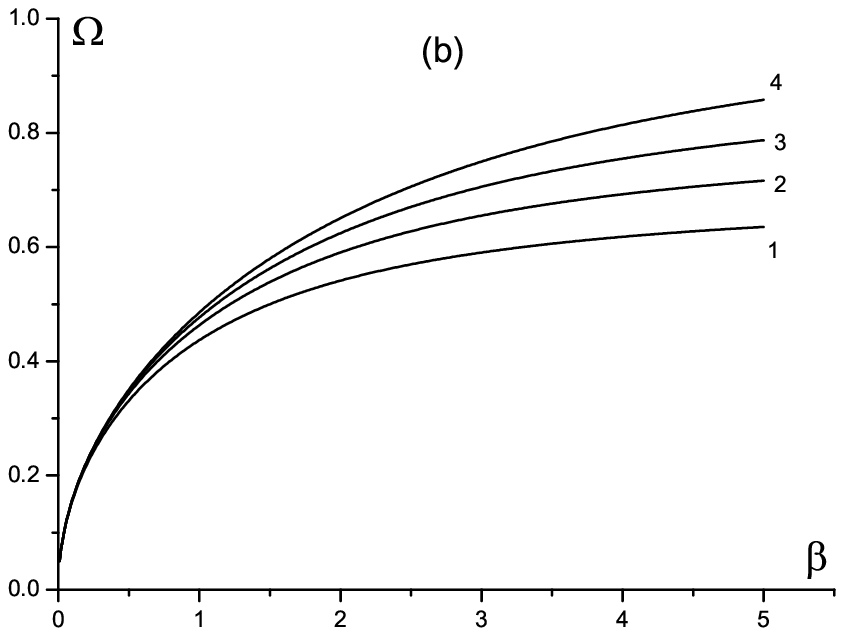}
\includegraphics[height=6.0cm]{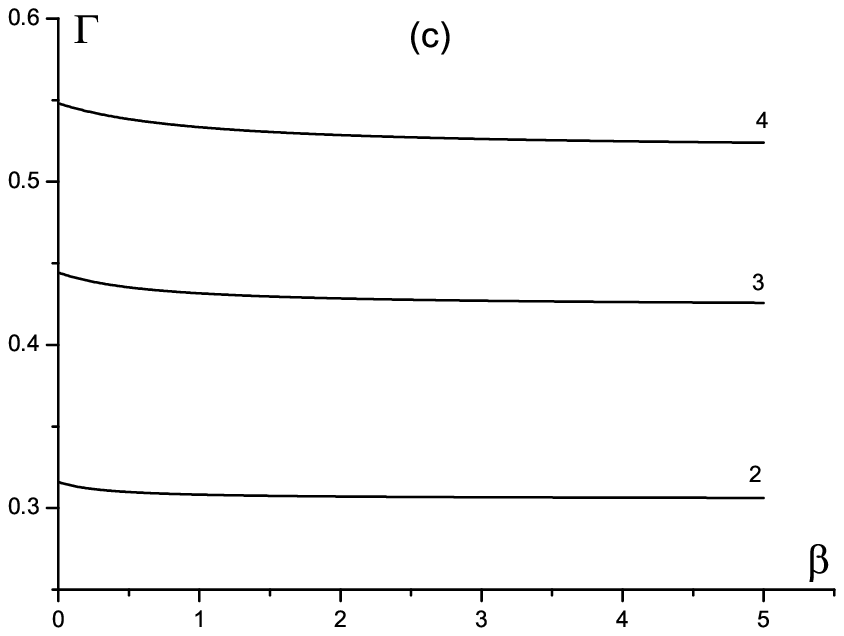}
\caption{ The frequency $\Omega$ of the fast wave (a) and the slow wave (b), and the 
growth rate $\Gamma$ of the unstable mode (c) versus $\beta$: $\delta=0,~0.02,~0.05,
~0.1$ from bottom to top;  $\kappa=10$, $\rr{cos}^2\theta=0.5$; as $\Gamma=0$ at $\delta=0$ curve 1 is not shown on pannel (c).}
\end{figure}

\end{document}